\documentclass{article}
\usepackage[utf8]{inputenc}
\usepackage{float}
\usepackage{jheppub}
\usepackage{amsmath}
\usepackage[english]{babel}
\usepackage{amssymb}
\usepackage{mathtools}
\usepackage{color}
\definecolor{ao}{rgb}{0.0, 0.5, 0.0}

\usepackage{graphicx}
\usepackage{longtable}
\usepackage{tikz}
\usepackage{comment}
\usepackage{capt-of}
\usepackage{amsthm}
\usepackage{hyperref}
\hypersetup{
    colorlinks=true,
    linkcolor=blue
    }

\usepackage{bm}

\usepackage{caption}
\usepackage{subcaption}

\newcommand{\SOMMA}[2]{\displaystyle\sum\limits_{#1}^{#2}}

\usepackage{tikz}
\usetikzlibrary{backgrounds}
\usepackage{listofitems} 
\usetikzlibrary{arrows.meta} 
\usepackage[outline]{contour} 
\contourlength{1.4pt}

\tikzset{>=latex} 
\usepackage{xcolor}

\title{Beyond Disorder: Unveiling Cooperativeness in Multidirectional Associative Memories}

\author[c,d]{Andrea Alessandrelli,}
\author[b,e]{Adriano Barra,}
\author[a]{Andrea Ladiana,}
\author[a]{Andrea Lepre,}
\author[g,h,i]{Federico Ricci-Tersenghi.}

\affiliation[a]{Dipartimento di Matematica e Fisica, Università del Salento, Lecce, Italy.}

\affiliation[b]{Istituto Nazionale d'Alta Matematica, GNFM, Roma, Italy.}

\affiliation[c]{Dipartimento di Informatica, Università di Pisa, Pisa Italy.}

\affiliation[d]{Istituto Nazionale di Fisica Nucleare, Sezione di Lecce, Italy.}

\affiliation[e]{Dipartimento di Scienze di Base Applicate all'Ingegneria, Sapienza Universit\`a di Roma, Rome, Italy.}

\affiliation[g]{Dipartimento di Fisica, Sapienza Universit\`a di Roma, Roma, Italy.}

\affiliation[h]{CNR-Nanotec, Rome unit, 00185 Roma, Italy.}

\affiliation[i]{Istituto Nazionale di Fisica Nucleare, Sezione di Roma1, Italy}

\abstract{By leveraging tools from the statistical mechanics of complex systems, in these short notes we extend the architecture of a neural network for hetero-associative memory called \textit{three-directional associative memories} (TAM) to explore supervised and unsupervised learning protocols. In particular, by providing entropic-heterogeneous datasets  to its various layers,   we predict and quantify a new emergent phenomenon  -that we term layer's ``cooperativeness"- where the interplay of dataset entropies across network's layers enhances their retrieval capabilities beyond those they would have without reciprocal influence. Naively we would expect layers trained with less informative datasets to develop smaller retrieval regions compared to those pertaining to layers that experienced more information: this does not happen and all the retrieval regions settle to the same amplitude, allowing for optimal retrieval performance globally. This cooperative dynamics marks a significant advancement in understanding emergent computational capabilities within disordered systems.
}

\begin{document}

\maketitle

\section{Introduction}

John Hopfield's legacy, recently recognized with the Nobel Prize in Physics, continues to inspire the study of associative memories. His groundbreaking work established the foundation for understanding how networks of simple units can give rise to complex, emergent behaviors. Thanks to modern variations of Hopfield's networks, they are experiencing a resurgence of interest \cite{krotov2023new,krotov2016dense,krotov2020large,ramsauer2020hopfield,barra2023thermodynamics,ExpHop1,LucibelloMezard,Lenka}.

Building on the layered associative Hebbian network architecture introduced for pattern recognition and disentanglement tasks \cite{TAM_easy, TAM_hard}, this paper extends the exploration to new learning paradigms, pushing the boundaries of what these associative networks may accomplish.
\\
While earlier work demonstrated how such networks could autonomously extract fundamental components from composite inputs—like identifying individual notes from musical chords—we delve deeper into the dynamics that govern these emergent capabilities. 
This dual framework enables a comprehensive examination of how varying levels of data structure and supervision influence the network's performance, especially under noisy or corrupted conditions \cite{SUPdense, UNSUPdense}.
\\
The most striking finding of our study is the emergence of a phenomenon we term ``cooperativeness". A detailed examination of the network phase diagrams, parameterized by dataset entropy values across distinct layers, reveals that the retrieval performance of each layer is not merely a reflection of its corresponding dataset's quality. Instead, it is governed by the collective interplay of datasets' entropy distributions across all layers. Interestingly, heterogeneous entropy levels diminish the retrieval performance of layers trained over more high quality datasets while enhancing the performance of those associated with noisier ones. This dynamics results in a balanced retrieval capacity across the network, a synergistic interaction absent in classical associative networks where layers operate independently. Notably, our approach leverages tools from the statistical mechanics of complex systems \cite{MPV,Talagrand,PRL-NOI,Fede0,Albe1,LindaRSB}, allowing us to rigorously predict and quantify this cooperative phenomenon through mathematically robust frameworks. This cooperative dynamics represents a significant advancement in understanding emergent intelligence in disordered systems.

\section{The Supervised and Unsupervised Hebbian protocols}
We consider a neural network composed of three different families of binary neurons hereafter indicated by $\bm\sigma\equiv\{\sigma^A_i\}^{A=1,2,3}_{i= 1, \dots, N_A}$  which interact in pairs via generalized Hebbian couplings (\textit{vide infra}) whose goal lies in reconstructing the information encoded in a triplet of $K$ binary archetypes $\{\bm\xi_\mu^A\}^{A=1,2,3}_{\mu=1, \dots, K}$ respectively of length $N_1, N_2$ and $N_3$.
However, such archetypes are not provided directly to the network, hence the latter has to infer them by experiencing solely their noisy or corrupted versions. In particular, we assume that for each triplet of archetypes $(\mu, A)$, $M$ examples $\eta_\mu^{a,A}$, with $a = 1, \hdots , M$ , are available, which are corrupted versions of the archetypes, such that for each $A=1,2,3$ and $i=1,\dots, N_A$ we have
\begin{equation}
	\mathbb{P}(\eta_{i,\mu}^{a,A}|\xi_{i,\mu}^A)= \dfrac{1+r_A}{2}\delta(\eta_{i,\mu}^{a,A}-\xi_{i,\mu}^{A})+\dfrac{1-r_A}{2}\delta(\eta_{i,\mu}^{a,A}+\xi_{i,\mu}^{A})
\end{equation}
where $r_A \in [0, 1]$ rules the quality of the dataset, i.e. for $r_A = 1$ the example matches perfectly the archetype, while for $r_A = 0 $ it is totally random. To quantify the information content of the dataset it is useful to introduce the variables
\begin{equation}
\label{eq:rho_def}
\begin{array}{llll}
        \rho_A = \dfrac{1-r_A^2}{M r_A^2}, \quad {\rho_{A B}}=\dfrac{1-r_{A}^{2}r_{B}^{2}}{M r_{A}^{2}r_{B}^{2}}\;\;\;\;\mathrm{with}\;\,\,A,B\in\{1,2,3 \}
\end{array}
\end{equation}
that we shall refer to as the \textit{dataset entropies} as deepened in \cite{alemanno2023supervised, aquaro4520891hebbian}. We observe that both $\rho_A$ and $\rho_{AB}$ approaches zero either when the examples perfectly match the archetypes (i.e., $r_A, r_B \to 1$), when the number of examples becomes infinite (i.e., $M \to \infty$), or under both conditions simultaneously.

The information related to the archetypes is encoded in the synaptic matrix, as outlined by the following cost function (or Hamiltonian):

\begin{figure}
    \centering
    \includegraphics[width=0.35\linewidth]{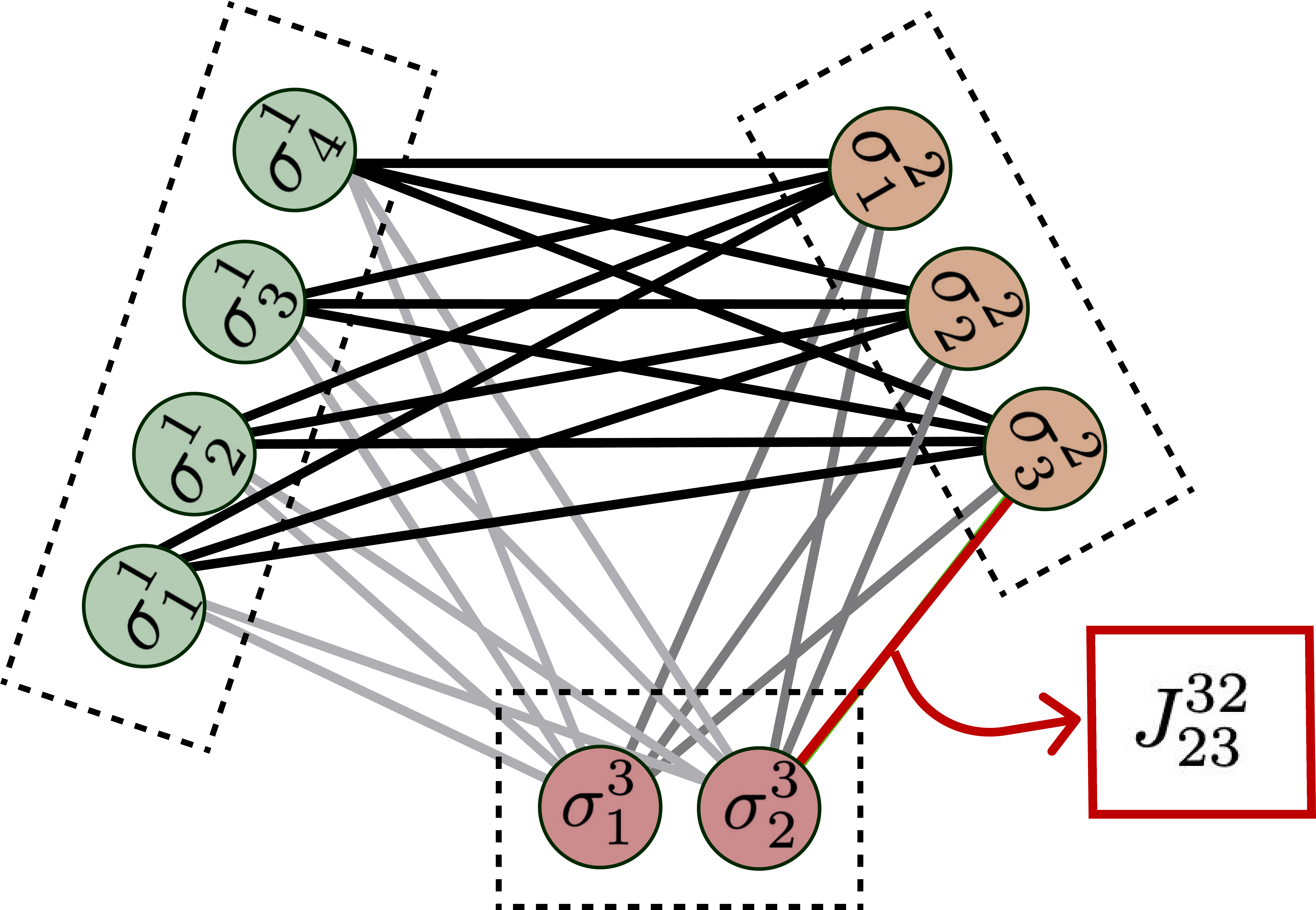}
    \caption{Schematic representation of the neural network described in \eqref{eq:H-TAM-generic} for the case $(N_1,N_2,N_3)=(4,3,2)$.}
    \label{fig:toy_model}
\end{figure}

\begin{equation}\label{eq:H-TAM-generic} 
\begin{array}{lll} \mathcal{H}^{\bm g}_{\bm N}(\bm\sigma|\bm J)=-\dfrac{1}{2}\SOMMA{A\neq B}{3}g_{AB}\SOMMA{i,j =1}{N_A, N_B} J_{ij}^{AB}\sigma_i^A\sigma_j^B, \end{array} \end{equation} where $\boldsymbol g \in \mathbb R^{3 \times 3}$ represents the strength of the interactions between different layers. The network architecture is sketched in Fig. \ref{fig:toy_model}. In order to let the network deal with examples rather than patterns, in this paper we inspect the following two variations of the above coupling matrix:

\begin{equation}\label{eq:H-TAM-unsup} \begin{array}{lll} (J^{(unsup)})_{ij}^{AB}= \dfrac{1}{r_A r_B \sqrt{N_A N_B(1+\rho_A)(1+\rho_B)}}\SOMMA{\mu=1}{K}\dfrac{1}{M}\SOMMA{a=1}{M}\eta_{i,\mu}^{a, A}\eta_{j,\mu}^{a, B}, \end{array} \end{equation}

\begin{equation}\label{eq:H-TAM-sup} \begin{array}{lll} (J^{(sup)})_{ij}^{AB}= \sqrt{\dfrac{(1+\rho_A)(1+\rho_B)}{N_A N_B}}\SOMMA{\mu=1}{K}\left(\dfrac{1}{M}\SOMMA{a=1}{M}\eta_{i,\mu}^{a, A}\right)\left(\dfrac{1}{M}\SOMMA{b=1}{M}\eta_{j,\mu}^{b, B}\right). \end{array} \end{equation}

In the first case, there is no external teacher who knows the labels and can organize the examples based on archetypes, as occurs in the second scenario. This distinction is why the two formulations are associated with unsupervised and supervised learning protocols, respectively \cite{SUPdense, UNSUPdense, alemanno2023supervised, aquaro4520891hebbian}.

As calculations will be performed in the thermodynamic limit, where $N_1, N_2, N_3\to\infty$, it is important to highlight that the sizes of the three layers—and consequently the lengths of the corresponding examples—as well as the number of samples in each dataset, can differ from one another, meaning  $N_1\neq N_2\neq N_3$ and $M_1\neq M_2\neq M_3$. Furthermore, despite these differences, the number of archetypes remains constant across all layers, denoted by $K$ for each.
Moreover, in order to ensure a meaningful asymptotic (thermodynamic) behavior, the ratio between the number of patterns and their respective lengths must remain finite. To achieve this, we impose the following conditions on $K$, $N_1$, $N_2$ and $N_3$:
\begin{equation}
    \begin{array}{lll}
         \lim\limits_{N_1,N_3\to\infty}\sqrt{\dfrac{N_1}{N_3}}=\alpha\,,   \quad\lim\limits_{N_1,N_2\to\infty}\sqrt{\dfrac{N_1}{N_2}}=\theta\,, \quad\lim\limits_{N_1,K\to\infty}\dfrac{K}{N_1}=\gamma
    \end{array}
\end{equation}
where $\alpha, \theta, \gamma \in \mathbb{R}^+$ are {\em control parameters}. The parameter $\gamma$ characterizes the storage capacity of the network, and our focus will be on the high-storage regime, where $\gamma >0$.

\begin{figure}[t]
    \centering
    \includegraphics[width=\linewidth]{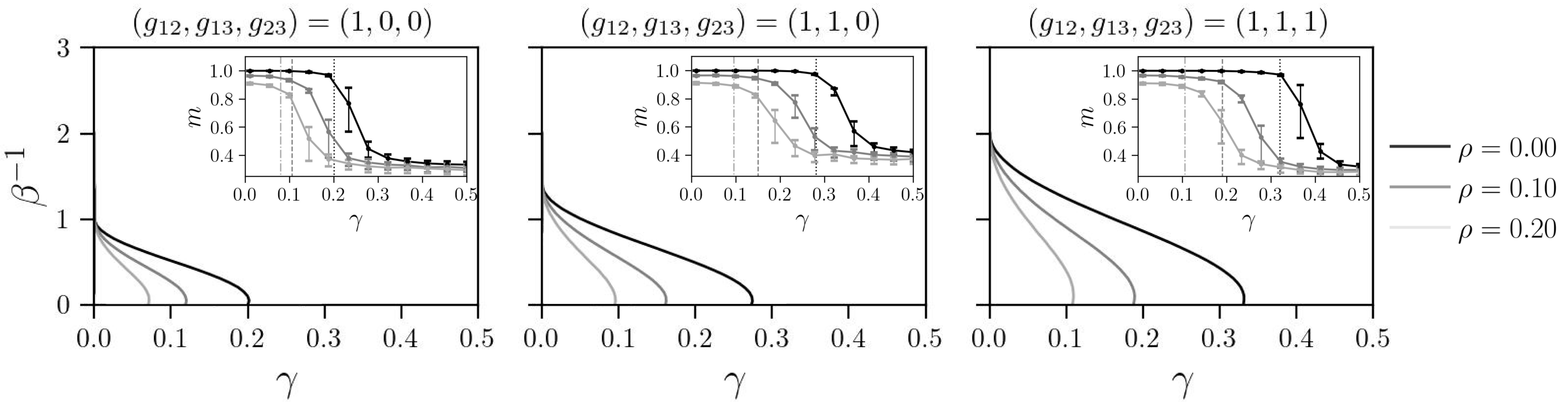}
    \caption{Phase diagrams of the TAM network in the supervised setting  in the noise versus storage plane at $\alpha = \theta = 1$. The analysis includes different inter-layer interaction strengths and various values of the dataset entropy $\rho$, as indicated in the legend. Each solid line depicts the phase transition for the whole network splitting the working region (bottom left) -where archetypes are learned and thus their retrieval and generalization allowed- from the blackout region (up right) -where spin glass effects prevail- for a specific value of dataset entropy i.e., $\rho_1=\rho_2=\rho_3=\rho$. The retrieval region is determined by the conditions $|m^1_{\xi_1^1}|, |m^2_{\xi_1^1}|, |m^3_{\xi_1^1}| > 0$: these inequalities are all satisfied simultaneously in the region below the solid line, while above it, all magnetizations vanish. The influence of $\rho$ is clearly visible: as $\rho$ increases, the retrieval region progressively shrinks in all diagrams. For $\rho = 0$, we recover the results of the standard Kosko's BAM case \cite{kosko} (first panel) and the novel ones pertaining to the TAM \cite{TAM_easy} (second and third panels). In the insets of each plots: MC simulation at zero-fast noise ($\beta^{-1} = 0$) with a symmetric network ($N_1 = N_2 = N_3 = 1000$), showing the evolution of the Mattis magnetizations $m_{\xi_1}$ across the layers as a function of network load ($\gamma$) for different $\rho$. The simulations agree with theoretical predictions, correctly depicting the maximum load beyond which the network stops functioning.}
        \label{fig:phaseDiagram}
\end{figure}

Pivotal for a statistical mechanical analysis is the study of the quenched free energy in the thermodynamic limit, defined as 
\begin{equation}\label{eq:Free-Definition}
\mathcal A_{\alpha,\theta,\gamma}^{\bm g}(\beta) = \lim_{N_1,N_2,N_3 \to \infty}\mathbb{E}\frac{1}{L}\ln \left[\SOMMA{\{\bm\sigma^1,\bm\sigma^2,\bm\sigma^3\}}{}\exp\left(-\beta \mathcal H^{\bm g}_{\bm N}(\bm \sigma|\bm J)\right)\right], 
\end{equation}
where $\mathbb{E}$ averages over the $\bm J$ distributions, $L = \frac{1}{3}\left(\frac{1}{\sqrt{N_1 N_2}}+\frac{1}{\sqrt{N_1 N_3}}+\frac{1}{\sqrt{N_2 N_3}}\right)$ and $\beta  \in \mathbb{R}^+$ tunes the fast noise in the network such that for $\beta \to 0^+$ network's dynamics is a pure random walk in the neural configuration space (and any configuration is equally likely to occur), while for $\beta\to+\infty$ its dynamics steepest descends toward the minima of the Hamiltonian \eqref{eq:H-TAM-generic}.
\newline
More precisely, our aim is to find an expression of $\mathcal A_{\alpha,\theta,\gamma}^{\bm g}(\beta)$ in terms of a suitable set of macroscopic observables able to capture the global behavior of the system: these \emph{order parameters} are the $K$ archetype (ground truth) Mattis magnetizations that assess the quality of network's retrieval, defined as 
\begin{eqnarray} \label{eq:ma}
         m_{\xi_\mu^A}^A &=& \dfrac{1}{N_A}\SOMMA{i=1}{N_A}\xi_{(i,\mu)}^A\sigma_i^A,
\end{eqnarray}
such that $m_{\xi_\mu^A}^A=1$ accounts for a perfect retrieval of the archetype $\xi^{\mu}$ by layer $A$, its lacking being accounted by $m_{\xi_\mu^A}^A=0$.

The application of Guerra’s interpolation method \cite{guerra} allows us to derive an explicit expression for the quenched free energy in the thermodynamic limit, in terms of the control parameters ($\beta, \gamma, \alpha, \theta$ and $\bm g$) and the order ones, under the assumption of replica symmetry.y. This assumption implies that, in the thermodynamic limit, the observables defined in \eqref{eq:ma} exhibit negligible fluctuations around their means. Once the quenched free energy is expressed in terms of the control and order parameters, we can proceed to extremize it with respect to the order parameters. This process results in a set of self-consistent equations, whose solutions describe the behavior of the order parameters as functions of the control ones. By analyzing these solutions, we can construct the phase diagram, identifying regions in the control parameter space where the network successfully learns the archetypes from the examples, retrieves them  and it is thus capable of generalization.

\begin{figure}[t]
    \centering
    \includegraphics[width=\linewidth]{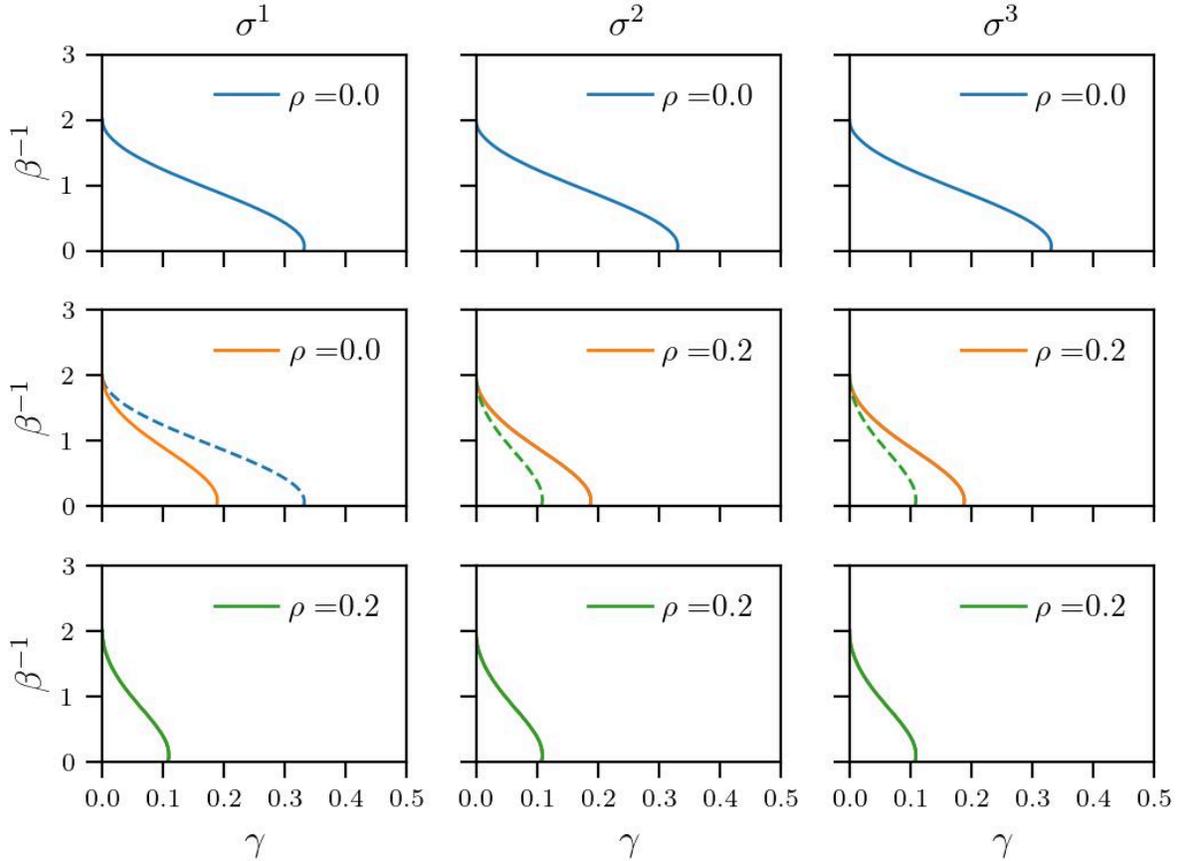}
    \caption{
    Phase diagrams in the symmetric case ($\alpha = \theta = 1$, $(g_{12}, g_{13}, g_{23}) = (1,1,1)$) highlighting the {\em cooperative behavior} among layers. Retrieval regions are shown for layers $\bm{\sigma}^1$, $\bm{\sigma}^2$, and $\bm{\sigma}^3$ under different dataset's entropy values: (top) zero-entropy datasets ($\rho_1 = \rho_2 = \rho_3 = 0$); (middle) heterogeneous-entropy datasets  ($\rho_2 = \rho_3 = 0.2$, $\rho_1 = 0$); (bottom) homogeneous-entropy datasets ($\rho_1 = \rho_2 = \rho_3 = 0.2$). The comparison highlights that the amplitude of the retrieval regions depends on the interplay between dataset entropies across layers: in the middle row, the first layer's retrieval region shrinks -orange curve- despite the noiseless inputs would allow to reach the dashed  blue line, allowing the noisy layers to expand theirs (from dashed green to orange boundary), an effect impossible without inter-layer interactions.  
}
    \label{fig:DifferentRho}
\end{figure}
Focusing specifically on the retrieval of the pattern triplet labeled by $\mu=1$ --without loss of generality-- we can extract an explicit expression for the self-consistency equations 
governing the order parameters, under the large dataset limit assumption (i.e. $M\gg 1$) which allow us to construct the model self diagrams both in the supervised and unsupervised scenarios.
\\
Focusing on the supervised protocol (i.e. assuming the network has the coupling \eqref{eq:H-TAM-sup}), we first investigate the case of datasets sharing the same entropy (i.e.,  \(\rho_1 = \rho_2 = \rho_3 = \rho\) over all the layers): results are summarized by the phase diagrams in the inter-layer activation strength \(\bm g\) vs noise presented in Fig. \ref{fig:phaseDiagram}\footnote{For the unsupervised counterpart, outcomes are qualitatively similar,  the only difference being in the definition of the dataset entropy (where \(\rho_A\) is replaced by \(\rho_{AB}\)).}. As shown in Fig. \ref{fig:phaseDiagram}, increasing \(\rho\) leads to a systematic reduction of the retrieval region—i.e., the domain in which the network can successfully reconstruct patterns from examples. As expected, for \(\rho = 0\) results collapse to those of the standard Hebbian-like TAM scenario \cite{TAM_easy}. The main reward of this analysis is the determination of the thresholds for learning, namely the minimal values of \(\rho\) required to sustain a non-zero retrieval region, as it allows us to predict, a priori, through \eqref{eq:rho_def}, the relationship between dataset quality (\(r_A\)) and dataset size (\(M\)) providing crucial insights for optimizing learning processes.

Then, by keeping the network symmetric in terms of both sizes and activation  strengths (i.e., $\alpha=\theta,\ g_{12}=g_{13}=g_{23}=1$), we deepened the analysis for datasets characterized by different entropies: results of this investigation are shown in Fig. \ref{fig:DifferentRho}. A detailed examination of the resulting phase diagrams, parameterized by the entropy values across the distinct layers, reveals an intriguing phenomenon: the retrieval region pertaining to each layer does not merely reflect the entropy of its corresponding dataset but is instead governed by the collective interplay of entropy distributions across all the  layers. Indeed, when the network has to handle entropic-heterogeneous datasets, a redistribution effect spontaneously appears: the retrieval region pertaining to the layer associated with the most informative dataset shrinks, while those of the layers at work with messy datasets enlarge, benefiting from their mutual interaction. This results in a more balanced retrieval performance across the network's layers. This emergent effect highlights their intrinsic cooperative nature, wherein the presence of a low entropy training set can enhance the performance of noisier ones, fostering a form of mutual reinforcement. 
We propose the term ``\emph{cooperativeness}" to describe this emergent property of the network, a feature that inherently arises from the reciprocal influence among the layers. This cooperative behavior is not merely a byproduct of parameter tuning but an intrinsic characteristic of the multipartite network structure, which only becomes evident through a comprehensive analytical treatment of the system’s self-consistency equations and whose presence can be crucial when dealing with dirty or small datasets.

\section{Conclusion}
Our work focuses on Hebbian information processing by a hetero-associative model able to cope with three sources of information simultaneously (i.e. the TAM network). In our setting, rather than directly providing the network with the original archetypes (i.e. the patterns), we expose it to examples—corrupted versions of them, thereby assessing its ability to learn and generalize from incomplete or noisy data.
\\
Through a statistical mechanics analysis, we obtained the phase diagrams of the network: the latter highlights how the amplitude of the retrieval region is affected by the entropy of the experienced datasets. Our results emphasize that successful pattern retrieval depends critically on both the quality and the quantity of examples provided, much like how human learning benefits from both clear instruction and repeated exposure.
\\
The most noteworthy finding of this study is the \emph{cooperative behavior} emerging among the layers of the network. A detailed examination of the phase diagrams, further corroborated by extensive numerical simulations, has revealed that layers associated with higher-informative datasets actively assist those provided with lower-informative ones, enhancing the amplitude of their retrieval regions by sacrificing their own. This effect arises because, the lower the entropy of a dataset, the larger the retrieval region of the relative layer  thus the stronger layer, benefiting from a higher quality dataset, can partially reduce its own retrieval region for the overall advantage of the system. This trade-off results in an optimal redistribution of learning and retrieval capacity across the network, fostering a form of mutual reinforcement that is absent in classical associative memory models.
\\
This phenomenon is particularly striking because it has no direct counterpart in the existing literature:  
our findings suggest that cooperativity may play a crucial and previously overlooked role in multidirectional associative models, potentially offering new insights into both artificial and biological memory systems and their applications.

\end{document}